\documentclass[english]{sbrt}
\IEEEoverridecommandlockouts
\usepackage{amsmath,amsfonts}
\usepackage[ruled,vlined]{algorithm2e}
\usepackage{array}
\usepackage{textcomp}
\usepackage{stfloats}
\usepackage{url}
\usepackage{verbatim}
\usepackage{graphicx}
\usepackage{cite}
\usepackage{xcolor}
\usepackage{subcaption}
\usepackage[nolist,nohyperlinks]{acronym} 
\usepackage{amsmath,amssymb,amsfonts}
\usepackage{algorithmic}
\usepackage{graphicx}
\usepackage{textcomp}
\usepackage{caption}
\usepackage{subcaption}
\usepackage{xspace}

\def\BibTeX{{\rm B\kern-.05em{\sc i\kern-.025em b}\kern-.08em
    T\kern-.1667em\lower.7ex\hbox{E}\kern-.125emX}}

\begin{document}
\begin{acronym}
	\acro{IRS}{Intelligent reflecting surface}
	\acro{RIS}{reconfigurable intelligent surface}
	\acro{irs}{intelligent reflecting surface}
	\acro{PARAFAC}{parallel factor}
	\acro{TALS}{trilinear alternating least squares}
	\acro{BALS}{bilinear alternating least squares}
	\acro{FA}{fluid antenna}
	\acro{FAS}{fluid antenna system}
	\acro{NTFAS}{Nested Tucker for Fluid Antenna Systems}
	\acro{DF}{decode-and-forward}
	\acro{AF}{amplify-and-forward}
	\acro{CE}{channel estimation}
	\acro{RF}{radio-frequency}
	\acro{THz}{Terahertz communication}
	\acro{EVD}{eigenvalue decomposition}
	\acro{CRB}{Cramér-Rao lower bound}
	\acro{CSI}{channel state information}
	\acro{BS}{base station}
	\acro{MIMO}{multiple-input multiple-output}
	\acro{NMSE}{normalized mean squared error}
	\acro{2G}{Second Generation}
	\acro{3G}{3$^\text{rd}$~Generation}
	\acro{3GPP}{3$^\text{rd}$~Generation Partnership Project}
	\acro{4G}{4$^\text{th}$~Generation}
	\acro{5G}{5$^\text{th}$~Generation}
	\acro{6G}{6$^\text{th}$~generation}
	\acro{E-TALS}{\textit{enhanced} TALS}
	\acro{UT}{user terminal}
	\acro{UTs}{users terminal}
	\acro{LS}{least squares}
	\acro{KRF}{Khatri-Rao factorization}
	\acro{KF}{Kronecker factorization}
	\acro{MU-MIMO}{multi-user multiple-input multiple-output}
	\acro{MU-MISO}{multi-user multiple-input single-output}
	\acro{MU}{multi-user}
	\acro{SER}{symbol error rate}
	\acro{SNR}{signal-to-noise ratio}
	\acro{SVD}{singular value decomposition}
\end{acronym}

\title{Joint Channel and Symbol Estimation for RIS-Assisted Fluid Antenna Systems}

\author{Josué V. de Araújo, Daniel C. Alcantara, Gilderlan T. de Araújo and André L. F. de Almeida
\thanks{Josué V. de Araújo, Daniel C. Alcantara and André L. F. de Almeida are with the Teleinformatics Department, Federal University of Ceará, Fortaleza-CE, e-mails: {danielchaves,josue.vas}@alu.ufc.br; andre@gtel.ufc.br.}%
\thanks{Gilderlan T. de Araújo is with the Federal Institute of Ceará, e-mail: gilerlan.tavares@ifce.edu.br.}
\thanks{This work is partially supported by the National Institute of Science and Technology (INCT-Signals) sponsored by Brazil’s National Council for Scientific and Technological Development (CNPq) (Proc. 406517/2022-3), and FUNCAP (Proc. INCT-25255-82587.32.41/64). The research of André L. F. de Almeida is partially supported by CNPq (Proc. 303356/2025-1). The research of Gilderlan T. de Araújo is supported by CNPq (Proc. 151870/2025-0).}
}

\maketitle


\renewcommand\baselinestretch{.96}

\begin{abstract}
This paper addresses joint channel and symbol estimation in reconfigurable intelligent surface (RIS)-aided multiuser uplink systems with fluid antennas (FAs) at the base station. We propose the Nested Tucker for Fluid Antenna Systems (NTFAS) protocol, in which FA port selection and user-dependent coding vary across blocks while the transmitted symbol matrix is shared across observations. This structure yields coupled Tucker models with common channel and data factors. A two-stage semi-blind bilinear alternating least squares (BALS) receiver is then developed to estimate the cascaded channel and symbols, and to separate the user-to-RIS and RIS-to-BS channels through the embedded PARAFAC structure. Simulations show that NTFAS improves cascaded-channel NMSE and spectral efficiency (SE) with respect to a competing semi-blind benchmark, while maintaining comparable BER performance.
\end{abstract}

\begin{keywords}
Fluid antenna system, reconfigurable intelligent surface, tensor decomposition, coupled Tucker model, semi-blind receiver, bilinear alternating least squares.
\end{keywords}

\renewcommand\baselinestretch{.88}

\section{Introduction}

Future sixth-generation (6G) wireless networks are expected to support stringent requirements on throughput, latency, reliability, connectivity, and energy efficiency \cite{Saad2020,Tataria2021}. Although massive multiple-input multiple-output (MIMO) and millimeter-wave massive MIMO have enabled large spatial gains, their scalability is constrained by hardware cost, power consumption, and the number of required RF chains \cite{Larsson2014,Busari2018}. RISs provide a complementary solution by shaping the propagation environment via passive reflecting elements, thereby improving coverage and link quality while consuming low power \cite{Wu2019,DiRenzo2020}.

Fluid antenna systems (FASs) provide another form of spatial adaptability by allowing the receiver to select favorable antenna ports within a compact aperture \cite{Wong2022,Wong2023}. This port-selection capability exploits small-scale spatial channel variations without requiring a separate RF chain for each candidate port. However, the same spatial reconfigurability that provides diversity also complicates channel acquisition, since the receiver observes different effective channels as the selected port changes.
Channel estimation for FAS and movable-antenna systems has received growing attention. Existing approaches include LMMSE estimation for large-scale FAS networks \cite{Ioannis_023_CE}, compressed-sensing-based methods for movable-antenna channels \cite{Ma_2023_CE}, and Bayesian reconstruction for FAS channels \cite{Zhang_2025}. These works demonstrate that channel acquisition is feasible by exploiting spatial correlation, sparsity, or prior channel structure. However, they primarily address stand-alone FAS/MA links with dedicated pilots and do not directly consider the cascaded RIS-FAS scenario, in which RIS phase shifts, FA port selections, user-dependent coding, and unknown multiuser symbols are jointly coupled in the received signal.

Combining RISs and FASs is attractive because the RIS reconfigures the propagation environment, while the FA receiver adapts its spatial sampling points. Existing RIS-aided movable/fluid antenna works have mainly focused on beamforming, antenna positioning, and sum-rate optimization under the assumption of available CSI \cite{Sun2025,Zhu2024a,Wei2025}. This assumption is restrictive, since RIS channel estimation already requires significant training overhead \cite{Zheng2020,Swindlehurst2022}, which becomes more critical when the effective channel also varies with FA port selection and multiuser transmission. This motivates a low-overhead receiver that exploits the coupled structure of the observations to jointly estimate the channels and the data.

Tensor-based signal processing provides a natural framework for exploiting multilinear structures in wireless communication systems \cite{Sidiropoulos2017}. It has been successfully applied to blind and semi-blind estimation in MIMO and space-time-frequency coded systems \cite{Almeida2007,Favier_TSP_2014}, as well as to tensor-based channel estimation in RIS-assisted links \cite{AraujoSAM2020,Araujo2021,ardah2021trice}. Building on our previous work on movable antenna systems \cite{Araujo_SBrT2025} and RIS-aided FAS receivers \cite{Araujo2026_Arxiv}, this paper addresses the above gap by proposing the NTFAS transmission strategy and a semi-blind receiver for joint channel and symbol estimation in RIS-aided multiuser FAS uplinks.\footnote{\textit{Notation}: Vectors are denoted by boldface lowercase letters ($\mathbf{a}$), matrices by boldface capital letters ($\mathbf{A}$), and tensors by calligraphic letters ($\boldsymbol{\mathcal{X}}$). The transpose and pseudo-inverse of $\mathbf{A}$ are denoted by $\mathbf{A}^T$ and $\mathbf{A}^\dagger$, respectively. The matrix $\mathbf{I}_N$ denotes the $N \times N$ identity matrix. The operator $D_i(\mathbf{A})$ forms a diagonal matrix from the $i$-th row of $\mathbf{A}$, while $\operatorname{diag}(\mathbf{a})$ forms a diagonal matrix from the vector $\mathbf{a}$. The operator $\operatorname{vec}(\cdot)$ vectorizes a matrix, whereas $\operatorname{unvec}_{I \times J}(\cdot)$ is its inverse. The symbols $\otimes$, $\diamond$, and $\oslash$ denote the Kronecker product, Khatri-Rao product, and element-wise division, respectively.}

The main contributions are summarized as follows:
\begin{itemize}
    \item We propose the NTFAS protocol, in which the FA port selection matrix and the user-dependent coding matrix vary across blocks, while the transmitted symbol matrix is shared by all observations.

    \item We show that the received signals can be represented as a set of coupled Tucker models sharing common symbol and cascaded-channel matrices.

    \item We derive a two-stage semi-blind receiver based on BALS. The first stage jointly estimates the transmitted symbols and the cascaded RIS-FA channel, whereas the second one exploits the embedded PARAFAC structure to separate the user-to-RIS and RIS-to-BS channels.

    \item We use the estimated CSI to perform joint FA port selection and RIS phase-shift optimization, and evaluate the resulting BER, NMSE, and SE performances against a competing benchmark previously introduced in \cite{Araujo2026_Arxiv}.
\end{itemize}

In this paper, we make use of the following identities:
\begin{align}
    \operatorname{vec}(\mathbf{A}\mathbf{B}\mathbf{C}) 
    &= 
    (\mathbf{C}^T \otimes \mathbf{A})\operatorname{vec}(\mathbf{B}), 
    \label{eq:prop_vec_kron} \\
    (\mathbf{A} \otimes \mathbf{B})(\mathbf{C} \diamond \mathbf{D}) 
    &= 
    (\mathbf{AC} \diamond \mathbf{BD}), 
    \label{eq:prop_mixed_kr} \\
    \operatorname{vec}(\mathbf{A}\operatorname{diag}(\mathbf{x})\mathbf{B}) 
    &= 
    (\mathbf{B}^T \diamond \mathbf{A})\mathbf{x}. 
    \label{eq:prop_vec_diag}
\end{align}

\begin{figure}[!t]
    \centering
    \begin{minipage}[b]{0.47\textwidth}
        \centering
        \includegraphics[width=\textwidth]{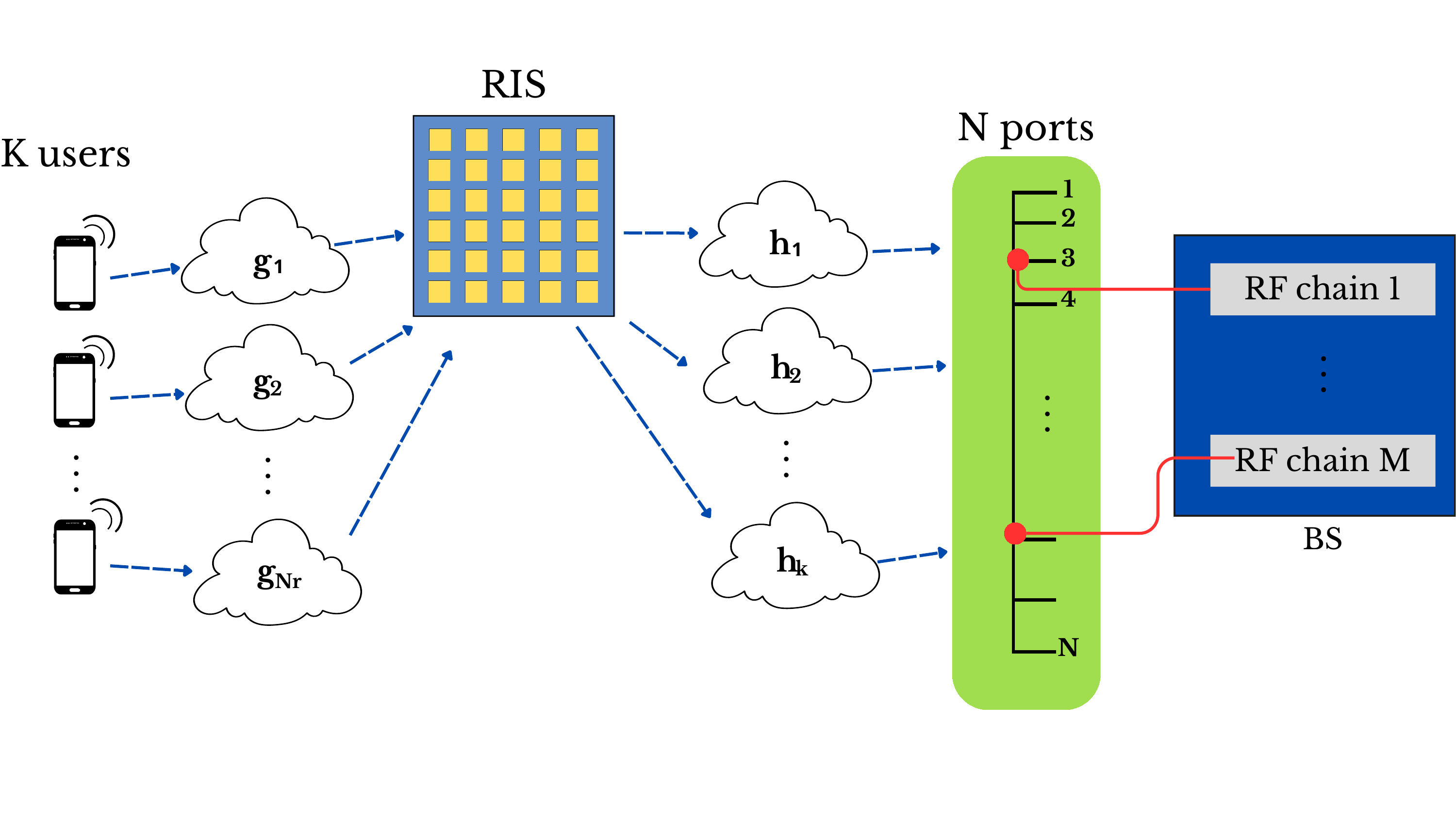}
        \caption{\small{RIS-assisted FA-based uplink \ac{MU} scenario.}}
        \label{fig:system}
    \end{minipage}
    \hfill
    \begin{minipage}[b]{0.47\textwidth}
        \centering
        \includegraphics[width=\textwidth]{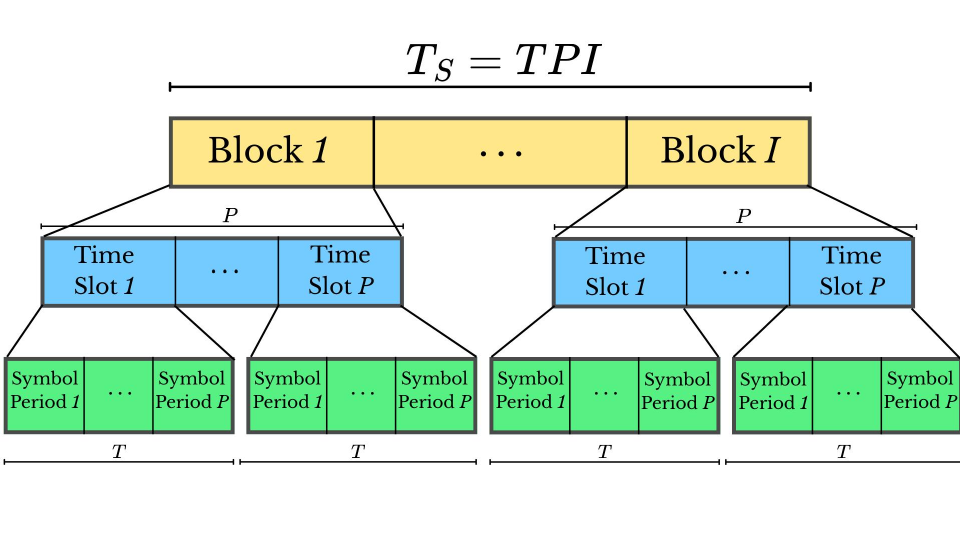}
        \caption{Transmission time structure.}
        \label{fig:time}
    \end{minipage}
    \label{fig:geral}
\end{figure}

\section{System Model and Tensor Formulation}

We consider a multiuser uplink communication system comprising $K$ single-antenna users, an RIS with $N_r$ elements, and a base station equipped with $M$ active RF chains connected to $N$ FA ports, similarly to the architecture described in \cite{Araujo_SBrT2025}. To significantly enrich temporal diversity and improve channel identifiability, we propose NTFAS. In this hierarchical frame structure, the FA port selection matrix $\mathbf{S}_i \in \{0,1\}^{M \times N}$ and the temporal coding matrix $D_i(\mathbf{C}) \in \mathbb{C}^{K \times K}$ are fixed during the $i$-th block. Conversely, the RIS phase-shift matrix $D_p(\mathbf{\Theta}) \in \mathbb{C}^{N_r \times N_r}$ varies dynamically at each $p$-th time slot. Furthermore, the transmitted symbol matrix $\mathbf{X} \in \mathbb{C}^{K \times T}$ remains invariant across the $I$ blocks, allowing the same data block to be observed under multiple FA selection and temporal coding configurations. The $(i,p)$-th component of the received signal is formulated as
\begin{equation}
    \mathbf{Y}_{i,p}
    =
    \mathbf{S}_i \mathbf{H} D_p(\mathbf{\Theta}) \mathbf{G} D_i(\mathbf{C}) \mathbf{X}
    +
    \mathbf{Z}_{i,p}
    \in \mathbb{C}^{M \times T}.
\end{equation}

Vectorizing this expression while using property \eqref{eq:prop_vec_diag} yields
\begin{equation}
    \mathbf{y}_{i,p}
    =
    \left(
    (\mathbf{G} D_i(\mathbf{C}) \mathbf{X})^T
    \diamond
    \mathbf{S}_i \mathbf{H}
    \right)
    \boldsymbol{\theta}_p
    +
    \mathbf{z}_{i,p}
    \in \mathbb{C}^{TM \times 1},
\end{equation}
where $\mathbf{y}_{i,p} \triangleq \operatorname{vec}(\mathbf{Y}_{i,p})$, $\boldsymbol{\theta}_p \in \mathbb{C}^{N_r \times 1}$, and $\mathbf{z}_{i,p} = \operatorname{vec}(\mathbf{Z}_{i,p})$. By concatenating these vectorized observations across all $P$ time slots, we construct the aggregated signal matrix $\mathbf{Y}_i \triangleq [\mathbf{y}_{i,1}, \dots, \mathbf{y}_{i,P}]$ for the $i$-th block:
\begin{equation}
    \mathbf{Y}_i
    =
    \left(
    (\mathbf{G} D_i(\mathbf{C}) \mathbf{X})^T
    \diamond
    \mathbf{S}_i \mathbf{H}
    \right)
    \mathbf{\Theta}^T
    +
    \mathbf{Z}_i
    \in \mathbb{C}^{TM \times P},
    \label{EQ: receved Yi}
\end{equation}
where $\mathbf{\Theta} \in \mathbb{C}^{P \times N_r}$ collects the complete set of RIS phase shifts, and $\mathbf{Z}_i \in \mathbb{C}^{TM \times P}$ is the aggregated noise matrix. Using the property \eqref{eq:prop_mixed_kr}, the received signal in \ref{EQ: receved Yi} can be rewritten as
\begin{equation}
    \mathbf{Y}_i
    =
    (\mathbf{X}^T \otimes \mathbf{I}_M)
    (D_i(\mathbf{C}) \otimes \mathbf{S}_i)
    \boldsymbol{\Phi}^T
    +
    \mathbf{Z}_i
    \in \mathbb{C}^{TM \times P},
\end{equation}
where
\begin{equation}
    \boldsymbol{\Phi}
    \triangleq
    \mathbf{\Theta}
    (\mathbf{G}^T \diamond \mathbf{H})^T
    \in \mathbb{C}^{P \times KN}
    \label{eq:cascaded}
\end{equation}
denotes the composite cascaded channel matrix.

Finally, to fully decouple the symbol matrix $\mathbf{X}$ from the spatial variables, we define a third-order core tensor $\boldsymbol{\mathcal{A}}_i \in \mathbb{C}^{KN \times M \times K}$, whose mode-1 unfolding encapsulates the combined coding and port selection terms, i.e., $[\boldsymbol{\mathcal{A}}_i]_{(1)}=D_i(\mathbf{C}) \otimes \mathbf{S}_i^T,
    \, i=1,\ldots,I$.
In tensor notation, the noiseless observation tensor $\overline{\boldsymbol{\mathcal{Y}}}_i \in \mathbb{C}^{P \times M \times T}$ is compactly formulated as
\begin{equation}\label{eq:t3_data}
    \overline{\boldsymbol{\mathcal{Y}}}_i
    =
    \boldsymbol{\mathcal{A}}_i
    \times_1
    \boldsymbol{\Phi}
    \times_2
    \mathbf{I}_M
    \times_3
    \mathbf{X}^T.
\end{equation}
By applying standard tensor matricization rules, the mode-1 unfolding directly isolates the cascaded channel matrix $\boldsymbol{\Phi}$:
\begin{equation}
    [\overline{\boldsymbol{\mathcal{Y}}}_i]_{(1)}
    =
    \boldsymbol{\Phi}
    [\boldsymbol{\mathcal{A}}_i]_{(1)}
    (\mathbf{X}^T \otimes \mathbf{I}_M)^T
    \in
    \mathbb{C}^{P \times MT}.
    \label{eq:mode1Y}
\end{equation}
while the mode-3 unfolding decouples the symbol matrix $\mathbf{X}$:
\begin{equation}
    [\overline{\boldsymbol{\mathcal{Y}}}_i]_{(3)}
    =
    \mathbf{X}^T
    [\boldsymbol{\mathcal{A}}_i]_{(3)}
    (\mathbf{I}_M \otimes \boldsymbol{\Phi})^T
    \in
    \mathbb{C}^{T \times PM}.
    \label{eq:mode3Y}
\end{equation}
These structured unfoldings serve as the basis for the subsequent estimation stages of the ALS algorithm \cite{comon_2009}.

\section{Proposed Two-Stage Semi-Blind Receiver}
In the first stage, the Tucker structure is exploited to jointly estimate the global cascaded channel matrix $\boldsymbol{\Phi}$ and the common symbol matrix $\mathbf{X}$. In the second stage, the embedded PARAFAC structure of $\boldsymbol{\Phi}$ is used to decouple the individual channel matrices $\mathbf{G}$ and $\mathbf{H}$.

\subsection{First Stage: BALS Estimation of $\boldsymbol{\Phi}$ and $\mathbf{X}$}

By exploiting the mode-1 unfolding in \eqref{eq:mode1Y}, the global channel is estimated by solving the following LS problem:
\begin{equation}
    \hat{\boldsymbol{\Phi}}
    =
    \arg\min_{\boldsymbol{\Phi}}
    \sum_{i=1}^{I}
    \left\|
    [\boldsymbol{\mathcal{Y}}_i]_{(1)}
    -
    \boldsymbol{\Phi}
    \mathbf{V}_i
    \right\|_F^2,
\end{equation}
where $    \mathbf{V}_i
    \triangleq
    [\boldsymbol{\mathcal{A}}_i]_{(1)}
    (\mathbf{X}^T \otimes \mathbf{I}_M)^T
    =
    [\boldsymbol{\mathcal{A}}_i]_{(1)}
    (\mathbf{X} \otimes \mathbf{I}_M)
    \in
    \mathbb{C}^{KN \times MT}$.
The global problem can be compactly written as
\begin{equation}
    \hat{\boldsymbol{\Phi}}
    =
    \arg\min_{\boldsymbol{\Phi}}
    \left\|
    \mathbf{Y}_{(1)}
    -
    \boldsymbol{\Phi}
    \mathbf{V}
    \right\|_F^2,
\end{equation}
where $    \mathbf{Y}_{(1)}
    \triangleq
    \begin{bmatrix}
        [\boldsymbol{\mathcal{Y}}_1]_{(1)}
        &
        \cdots
        &
        [\boldsymbol{\mathcal{Y}}_I]_{(1)}
    \end{bmatrix}
    \in
    \mathbb{C}^{P \times IMT}$,
and $    \mathbf{V}
    \triangleq
    \begin{bmatrix}
        \mathbf{V}_1
        &
        \cdots
        &
        \mathbf{V}_I
    \end{bmatrix}
    \in
    \mathbb{C}^{KN \times IMT}$.
The estimate for the cascaded channel is obtained as
\begin{equation}
    \hat{\boldsymbol{\Phi}}
    =
    \mathbf{Y}_{(1)}
    \mathbf{V}^{\dagger}.
\end{equation}

From \eqref{eq:mode3Y}, $\mathbf{X}$ is estimated by solving the following problem:
\begin{equation}
    \hat{\mathbf{X}}^T
    =
    \arg\min_{\mathbf{X}^T}
    \sum_{i=1}^{I}
    \left\|
    [\boldsymbol{\mathcal{Y}}_i]_{(3)}
    -
    \mathbf{X}^T
    \mathbf{U}_i
    \right\|_F^2,
\end{equation}
where $    \mathbf{U}_i
    \triangleq
    [\boldsymbol{\mathcal{A}}_i]_{(3)}
    (\mathbf{I}_M \otimes \boldsymbol{\Phi})^T
    \in
    \mathbb{C}^{K \times PM}$.
 By horizontally concatenating the $3$-mode unfoldings and the corresponding regression matrices, we obtain
\begin{equation}
    \mathbf{Y}_{(3)}
    \triangleq
    \begin{bmatrix}
        [\boldsymbol{\mathcal{Y}}_1]_{(3)}
        &
        \cdots
        &
        [\boldsymbol{\mathcal{Y}}_I]_{(3)}
    \end{bmatrix}
    \in
    \mathbb{C}^{T \times IPM},
\end{equation}
and $    \mathbf{U}
    \triangleq
    \begin{bmatrix}
        \mathbf{U}_1
        &
        \cdots
        &
        \mathbf{U}_I
    \end{bmatrix}
    \in
    \mathbb{C}^{K \times IPM}$,
then $\mathbf{X}$ is found as
\begin{equation}
    \hat{\mathbf{X}}^T
    =
    \mathbf{Y}_{(3)}
    \mathbf{U}^{\dagger}.
\end{equation}

\subsection{Second Stage: BALS Estimation of $\mathbf{G}$ and $\mathbf{H}$}

The first stage provides the cascaded-channel estimate $\hat{\boldsymbol{\Phi}} \in \mathbb{C}^{P \times KN}$, whose transpose follows the PARAFAC structure, as follows
\begin{equation}
    \boldsymbol{\Phi}^T
    \approx
    (\mathbf{G}^T \diamond \mathbf{H})
    \mathbf{\Theta}^T
    \in
    \mathbb{C}^{KN \times P},
\end{equation}
such that $    \hat{\boldsymbol{\Phi}}_p
    =
    \operatorname{unvec}_{N \times K}
    (\hat{\boldsymbol{\phi}}_p)
    =
    \mathbf{H}
    D_p(\mathbf{\Theta})
    \mathbf{G}
    \in
    \mathbb{C}^{N \times K}$.

\subsubsection{Estimation of the RIS-to-BS Channel $\mathbf{H}$}

Let us define  $\mathbf{R}_p \triangleq D_p(\mathbf{\Theta}) \mathbf{G} \in \mathbb{C}^{N_r \times K}$. 
 By horizontally concatenating the local observations across all $P$ time slots, we have
\begin{equation}
    \underbrace{
    \begin{bmatrix}
        \hat{\boldsymbol{\Phi}}_1
        &
        \cdots
        &
        \hat{\boldsymbol{\Phi}}_P
    \end{bmatrix}
    }_{\hat{\boldsymbol{\Phi}}_{(H)}}
    =
    \mathbf{H}
    \underbrace{
    \begin{bmatrix}
        \mathbf{R}_1
        &
        \cdots
        &
        \mathbf{R}_P
    \end{bmatrix}
    }_{\mathbf{A}_H},
\end{equation}
then the RIS-to-BS channel can be estimated as
\begin{equation}
    \hat{\mathbf{H}}
    =
    \hat{\boldsymbol{\Phi}}_{(H)}
    \mathbf{A}_H^{\dagger}.
\end{equation}

\subsubsection{Estimation of the User-to-RIS Channel $\mathbf{G}$}

Let us define $\mathbf{Q}_p \triangleq D_p(\mathbf{\Theta}) \mathbf{H}^T \in \mathbb{C}^{N_r \times N}$, we establish the parallel concatenated system
\begin{equation}
    \underbrace{
    \begin{bmatrix}
        \hat{\boldsymbol{\Phi}}_1^T
        &
        \cdots
        &
        \hat{\boldsymbol{\Phi}}_P^T
    \end{bmatrix}
    }_{\hat{\boldsymbol{\Phi}}_{(G)}}
    =
    \mathbf{G}^T
    \underbrace{
    \begin{bmatrix}
        \mathbf{Q}_1
        &
        \cdots
        &
        \mathbf{Q}_P
    \end{bmatrix}
    }_{\mathbf{A}_G},
\end{equation}
which gives the following estimate for the User-to-RIS channel
\begin{equation}
    \hat{\mathbf{G}}^T
    =
    \hat{\boldsymbol{\Phi}}_{(G)}
    \mathbf{A}_G^{\dagger}.
\end{equation}

\section{Joint FA Port Selection and RIS Optimization}

After estimating $\hat{\mathbf{H}}$ and $\hat{\mathbf{G}}$, we use the estimated CSI to jointly select the FA ports and design the RIS phase shifts. For a given FA selection matrix $\mathbf{S}$ and RIS phase vector $\boldsymbol{\theta}=[\theta_1,\ldots,\theta_{N_r}]^T$, the estimated equivalent channel is
\begin{equation}
    \hat{\mathbf{H}}_{\mathrm{eq}}(\mathbf{S},\boldsymbol{\theta})
    =
    \mathbf{S}\hat{\mathbf{H}}
    \operatorname{diag}(\boldsymbol{\theta})
    \hat{\mathbf{G}}
    \in \mathbb{C}^{M \times K}.
\end{equation}
Assuming equal power allocation, the corresponding SE is
\begin{equation}
    R(\mathbf{S},\boldsymbol{\theta})
    =
    \log_2
    \det
    \left(
    \mathbf{I}_M
    +
    \frac{\rho}{K}
    \hat{\mathbf{H}}_{\mathrm{eq}}
    \hat{\mathbf{H}}_{\mathrm{eq}}^H
    \right),
\end{equation}
where $\rho$ denotes the SNR. The joint design problem is therefore
\begin{equation}
\begin{aligned}
    (\mathbf{S}^{\star},\boldsymbol{\theta}^{\star})
    &=
    \underset{\mathbf{S}\in\mathcal{S},\,\boldsymbol{\theta}}{\arg\max}
    \quad
    R(\mathbf{S},\boldsymbol{\theta}) \\
    &\text{s.t.}
    \quad
    |\theta_{n_r}|=1,\quad n_r=1,\ldots,N_r,
\end{aligned}
\end{equation}
where $\mathcal{S}$ contains all $\binom{N}{M}$ valid FA port selection matrices.

Due to the combinatorial search over $\mathcal{S}$ and the unit-modulus RIS constraint, we adopt an exhaustive search over the FA port combinations combined with an SVD-based RIS phase design. For each candidate $\mathbf{S}\in\mathcal{S}$, let $\bar{\mathbf{H}}=\mathbf{S}\hat{\mathbf{H}}$. The equivalent channel can be written as
\begin{equation}
    \hat{\mathbf{H}}_{\mathrm{eq}}
    =
    \sum_{n_r=1}^{N_r}
    \theta_{n_r}
    \bar{\mathbf{h}}_{n_r}
    \hat{\mathbf{g}}_{n_r,:},
\end{equation}
where $\bar{\mathbf{h}}_{n_r}$ is the $n_r$-th column of $\bar{\mathbf{H}}$ and $\hat{\mathbf{g}}_{n_r,:}$ is the $n_r$-th row of $\hat{\mathbf{G}}$. We form the tensor $\boldsymbol{\mathcal{Z}}\in\mathbb{C}^{M\times K\times N_r}$ with slices
\begin{equation}
    \mathbf{Z}_{n_r}
    =
    \bar{\mathbf{h}}_{n_r}
    \hat{\mathbf{g}}_{n_r,:}.
\end{equation}
Then, from the dominant left singular vector $\mathbf{u}_{3,1}$ of the mode-3 unfolding $[\boldsymbol{\mathcal{Z}}]_{(3)}$, the RIS phase vector is chosen as
\begin{equation}
    \boldsymbol{\theta}(\mathbf{S})
    =
    \exp
    \left(
    -j\angle \mathbf{u}_{3,1}
    \right).
\end{equation}
Finally, the pair $(\mathbf{S}^{\star},\boldsymbol{\theta}^{\star})$ is selected as the candidate that provides the largest estimated SE.

\begin{algorithm}[!t]
\footnotesize
\DontPrintSemicolon
\caption{Proposed two-stage BALS semi-blind receiver}
\label{alg:BALS_P3}
\KwIn{$\{\boldsymbol{\mathcal{Y}}_i\}_{i=1}^{I}$, $\{\boldsymbol{\mathcal{A}}_i\}_{i=1}^{I}$, $\boldsymbol{\Theta}$, ref. column $\mathbf{x}_1$, tolerance $\delta$}
\KwOut{$\hat{\mathbf{H}}$, $\hat{\mathbf{G}}$, and $\hat{\mathbf{X}}$}
Initialize $\hat{\mathbf{X}}$ randomly; set $\ell=0$, $\epsilon_{\mathrm{old}}=\infty$, $\Delta_{\epsilon}=\infty$
\While{$\Delta_{\epsilon} \geq \delta$}{
    Build $\mathbf{V}=[\mathbf{V}_1,\ldots,\mathbf{V}_I]$ from $\hat{\mathbf{X}}$ and $\{\boldsymbol{\mathcal{A}}_i\}_{i=1}^{I}$
    
    Update $\hat{\boldsymbol{\Phi}} \leftarrow \mathbf{Y}_{(1)}\mathbf{V}^{\dagger}$
    
    Build $\mathbf{U}=[\mathbf{U}_1,\ldots,\mathbf{U}_I]$ from $\hat{\boldsymbol{\Phi}}$ and $\{\boldsymbol{\mathcal{A}}_i\}_{i=1}^{I}$
    
    Update $\hat{\mathbf{X}}^T \leftarrow \mathbf{Y}_{(3)}\mathbf{U}^{\dagger}$
    
    Compute $\epsilon_{\mathrm{new}}$; set $\Delta_{\epsilon}=|\epsilon_{\mathrm{new}}-\epsilon_{\mathrm{old}}|$, $\epsilon_{\mathrm{old}}\leftarrow\epsilon_{\mathrm{new}}$, $\ell\leftarrow\ell+1$
}
$\mathbf{D}_{\alpha}\leftarrow\operatorname{diag}(\mathbf{x}_1\oslash\hat{\mathbf{x}}_1)$

Correct scaling: $\hat{\mathbf{X}}\leftarrow\mathbf{D}_{\alpha}\hat{\mathbf{X}}$, $\hat{\boldsymbol{\Phi}}_p\leftarrow\hat{\boldsymbol{\Phi}}_p\mathbf{D}_{\alpha}^{-1}$, $p=1,\ldots,P$

Second stage: $\hat{\mathbf{H}}\leftarrow\hat{\boldsymbol{\Phi}}_{(H)}\mathbf{A}_{H}^{\dagger}$, $\hat{\mathbf{G}}^T\leftarrow\hat{\boldsymbol{\Phi}}_{(G)}\mathbf{A}_{G}^{\dagger}$
\end{algorithm}

\section{Identifiability and Computational Complexity}
Note that the uniqueness in the LS sense of the proposed two-stage BALS receiver is directly related to the pseudo-inverse operations used to estimate $\boldsymbol{\Phi}$, $\mathbf{X}$, $\mathbf{H}$, and $\mathbf{G}$. From the updates of $\boldsymbol{\Phi}$ and $\mathbf{X}$, the corresponding regression matrices must satisfy $IMT \geq KN$ and $IPM \geq K$, respectively. Moreover, the second-stage BALS decoupling of the cascaded channel requires $PK \geq N_r$ and $PN \geq N_r$ for the LS updates of $\mathbf{H}$ and $\mathbf{G}$, respectively. These inequalities provide practical constraints for the system parameters. For instance, $K \leq \min\left(\frac{IMT}{N},IPM\right)$ and $N_r \leq P\min(K,N)$ guide the choice of the number of users, FA ports, RIS elements, blocks, and time slots to ensure well-posed LS updates.

Regarding computational complexity, the dominant cost per iteration of \textbf{Algorithm 1} comes from the SVD calculations required by the pseudo-inverse operations. Recalling that computing the SVD of a $P \times Q$ matrix has complexity $\mathcal{O}(PQ\min(P,Q))$, the complexity of the proposed receiver is approximately $\mathcal{O}\left(IMT(KN)^2 + IPMK^2 + PKN_r^2 + PNN_r^2\right)$. The first two terms are associated with the BALS estimation of $\boldsymbol{\Phi}$ and $\mathbf{X}$, while the last two terms come from the second-stage BALS decoupling of $\boldsymbol{\Phi}$ into $\mathbf{H}$ and $\mathbf{G}$.




\section{Simulation Results}
\begin{figure*}[t]
    \centering
    \begin{subfigure}{0.32\textwidth}
        \centering
        \includegraphics[width=\linewidth]{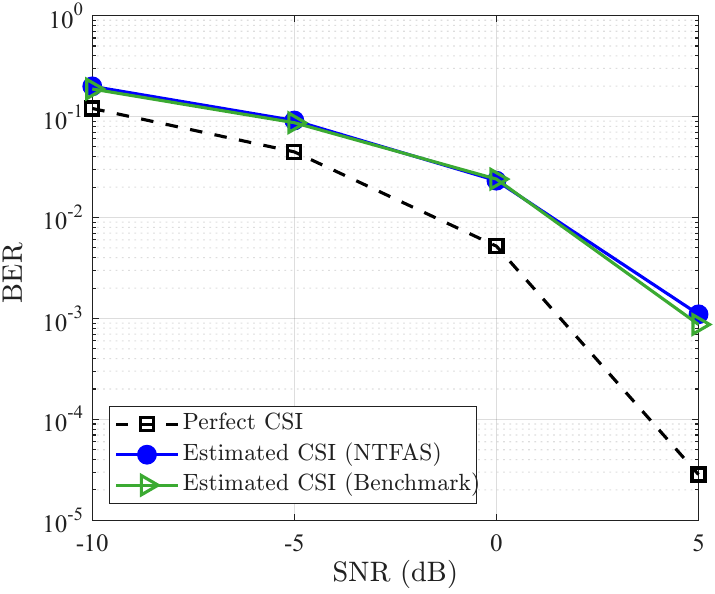}
        \caption{BER vs. SNR}
        \label{fig:ber}
    \end{subfigure}
    \hfill
    \begin{subfigure}{0.32\textwidth}
        \centering
        \includegraphics[width=\linewidth]{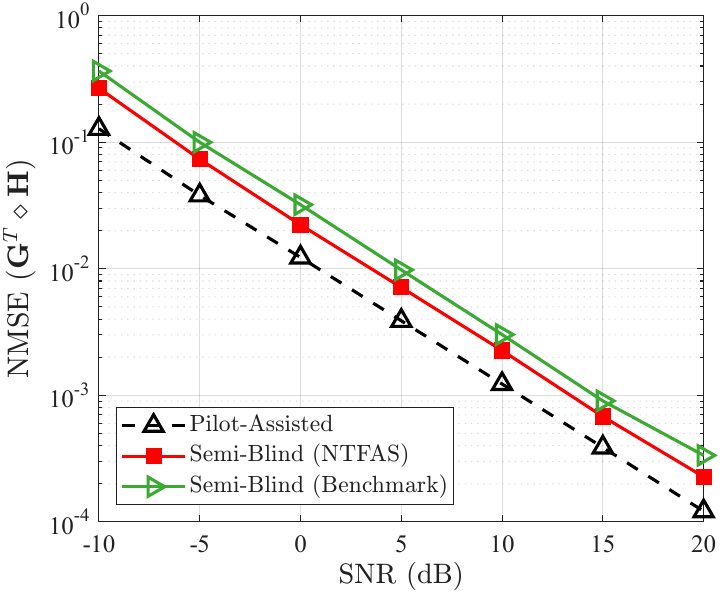}
        \caption{NMSE vs. SNR}
        \label{fig:nmse}
    \end{subfigure}
    \hfill
    \begin{subfigure}{0.32\textwidth}
        \centering
        \includegraphics[width=\linewidth]{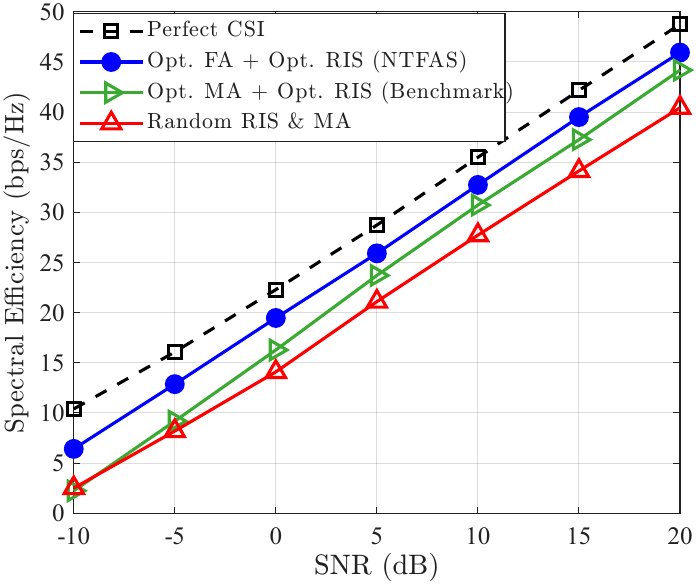}
        \caption{SE vs. SNR}
        \label{fig:se}
    \end{subfigure}
    \caption{Performance results versus SNR (dB): (a) BER, (b) NMSE, and (c) SE.}
    \label{fig:results}
\end{figure*}

\begin{figure}[t!]
    \centering
    \begin{subfigure}{0.4\textwidth}
        \centering
        \includegraphics[width=\linewidth]{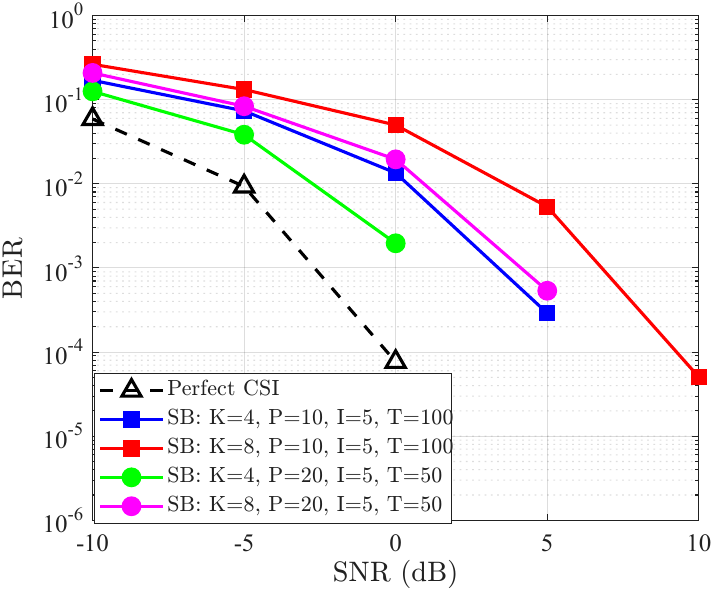}
        \caption{BER vs. SNR (dB)}
        \label{fig:ber_pkt}
    \end{subfigure}
    \hfill
    \begin{subfigure}{0.4\textwidth}
        \centering
        \includegraphics[width=\linewidth]{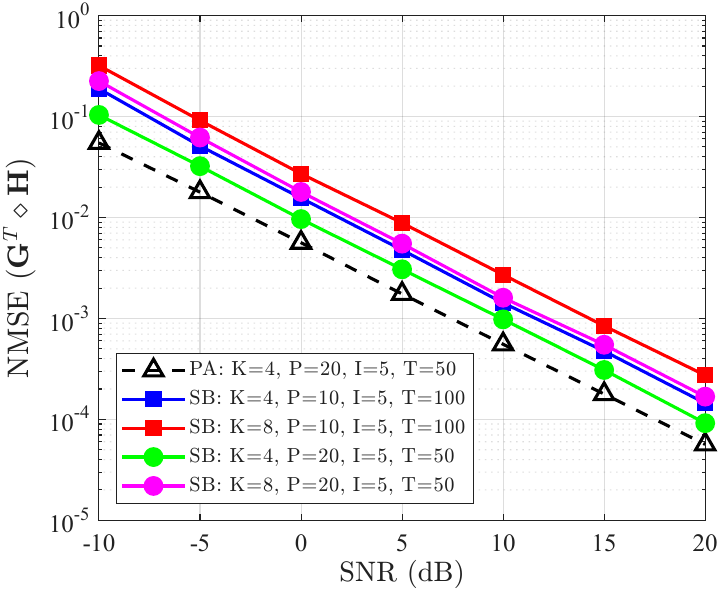}
        \caption{NMSE vs. SNR (dB)}
        \label{fig:nmse_pkt}
    \end{subfigure}
    \caption{Simulation results for the second scenario: (a) BER vs. SNR and (b) NMSE vs. SNR.}
    \label{fig:resuls_pkt}
\end{figure}
In this section, we assess the proposed semi-blind receiver for NTFAS in terms of bit error rate (BER), normalized mean square error (NMSE) of the cascaded channel factor $\mathbf{G}^{T}\diamond\mathbf{H}$, and SE after joint FA port selection and RIS phase-shift optimization. A previously proposed semi-blind receiver from \cite{Araujo2026_Arxiv} is used as the competing benchmark under comparable transmission-overhead conditions. The Perfect CSI receiver provides a reference bound for BER and SE, whereas the Pilot-Assisted/Data-Aided estimator is used as the NMSE benchmark. All BER curves are obtained with 16-PSK modulation.

Fig. \ref{fig:results} compares NTFAS with the competing semi-blind benchmark. In Fig. \ref{fig:ber}, both semi-blind receivers exhibit nearly identical BER behavior over the simulated SNR range, and both remain separated from the Perfect CSI curve. This is consistent with the proposed model: after the reference-symbol scaling correction, the symbol estimates are primarily limited by noise and residual channel-estimation error. Hence, the moderate NMSE difference between NTFAS and the benchmark is not sufficient to yield a visible BER gain at the considered operating points.

The NMSE curves in Fig. \ref{fig:nmse} show a clearer distinction. NTFAS consistently improves the cascaded-channel estimate with respect to the competing benchmark, while the Pilot-Assisted/Data-Aided estimator remains the best reference. This behavior is expected because NTFAS uses multiple block observations with different FA selection and coding matrices but a common symbol matrix, thereby increasing the number of coupled equations available for estimating $\boldsymbol{\Phi}$. The gain is moderate rather than dramatic, which is consistent with the fact that both semi-blind methods exploit tensor structure and satisfy the LS identifiability conditions discussed previously.

The SE results in Fig. \ref{fig:se} follow the NMSE trend. Since FA/RIS optimization is performed using the estimated CSI, a more accurate cascaded-channel estimate of NTFAS yields higher SE than the competing benchmark and keeps the curve closer to the Perfect CSI bound. The remaining gap to Perfect CSI is also expected because the optimization stage uses estimated, rather than true, channel matrices. Both optimized semi-blind schemes outperform the random RIS/antenna selection baseline, confirming that the estimated CSI remains useful for link adaptation. This gain is achieved without increasing the dominant receiver cost: under the same system dimensions, the dominant NTFAS term scales as $\mathcal{O}(IMT(KN)^2)$, whereas the benchmark contains the larger channel-update term $\mathcal{O}(MTI N_r^2(K^2+N^2))$. Thus, NTFAS provides a more favorable performance--complexity trade-off, improving NMSE and SE while maintaining comparable BER.

Fig. \ref{fig:resuls_pkt} evaluates the impact of $P$, $K$, and $T$ while keeping the same block count $I=5$ and comparable temporal overhead. For $K=4$, increasing the number of RIS phase configurations from $P=10$ to $P=20$ improves both BER and NMSE, even though the number of symbols is reduced from $T=100$ to $T=50$. This confirms that for the proposed receiver, additional RIS-domain diversity can improve the conditioning of the cascaded channel estimation problem.

Increasing the number of users from $K=4$ to $K=8$ degrades both metrics, as expected due to the larger number of coupled user-dependent components and the more demanding LS condition $IMT \geq KN$. For $K=8$, using $P=20$ instead of $P=10$ partially mitigates the degradation in both BER and NMSE, but it does not fully restore the performance achieved with $K=4$. Overall, the curves support the expected trade-off: larger $P$ improves the robustness of the semi-blind estimation by adding RIS-domain observations, whereas larger $K$ makes the joint channel-and-symbol recovery problem more difficult.

\section{Conclusions}
This paper proposed a two-stage semi-blind receiver for joint channel and symbol estimation in RIS-aided multiuser FAS uplinks. By varying the FA port selection and user-dependent coding matrices across blocks while keeping the transmitted symbol matrix common, the received signals were formulated as coupled Tucker models with shared symbol and cascaded-channel factors. The estimated cascaded channel was then factorized through its embedded PARAFAC structure to recover the user-to-RIS and RIS-to-BS channels individually. The proposed NTFAS provides a moderate but consistent NMSE improvement over the competing benchmark, resulting in higher SE after FA/RIS optimization, while both semi-blind methods exhibit similar BER performance. Our simulations also confirm the expected trade-off between model diversity and estimation difficulty: increasing the number of RIS phase configurations improves the conditioning of the estimation problem, whereas increasing the number of users makes joint recovery more challenging. Overall, NTFAS reduces pilot dependence while providing CSI estimates useful for FA port selection and RIS phase-shift design. Future work will address lower-complexity optimization, hardware impairments, time-varying channels, and multi-antenna users.

\renewcommand\baselinestretch{.97}

\bibliographystyle{IEEEtran}
\bibliography{IEEEexample}

\end{document}